\documentstyle[aps,12pt]{revtex}
\begin{document}
\title{Excited states of a static dilute spherical Bose condensate in a
trap\footnote{Contributed paper for
Low Temperature Conference LT21, Prague, August, 1996, to appear in
Czechoslovak Journal of Physics}}
\author{Alexander L. Fetter}
\address{Departments of Physics and Applied Physics, Stanford University,
Stanford, CA 94305-4060}
\date{\today}
\maketitle
\begin{abstract}The Bogoliubov approximation is used to study the excited
states of a dilute gas of $N$ atomic
bosons trapped in an isotropic harmonic potential characterized by a
frequency $\omega_0$ and an oscillator
length $d_0 = \sqrt{\hbar/m\omega_0}$. The  self-consistent static Bose
condensate has macroscopic occupation
number
$N_0
\gg 1$, with nonuniform  spherical condensate density
$n_0(r)$; by assumption, the depletion of the condensate is small ($N'
\equiv N - N_0\ll N_0$).  The linearized
density fluctuation operator
$\hat
\rho'$ and velocity potential operator $\hat\Phi '$ satisfy coupled
equations that embody particle conservation
and Bernoulli's theorem.
 For each angular momentum
$l$, introduction of quasiparticle operators yields coupled eigenvalue
equations for the excited states; they
can be expressed   either in terms of   Bogoliubov
coherence amplitudes
$u_l(r)$ and
$v_l(r)$ that determine the appropriate linear combinations of particle
operators, or in terms of
hydrodynamic amplitudes
$\rho_l'(r)$ and $\Phi_l'(r)$.  The hydrodynamic picture suggests a simple
variational approximation for $l >
0$ that provides an upper bound for the lowest eigenvalue
$\omega_l$   and an estimate for the corresponding zero-temperature
occupation number $N_l'$;   both
expressions  closely resemble
 those  for a  uniform bulk Bose condensate.

\end{abstract}
\section{INTRODUCTION}
Recent experimental verification of Bose condensation in dilute confined
$^{87}$Rb \cite{And}
has stimulated  extensive theoretical research.  In the Bogoliubov
approximation, $N_0$ atoms occupy the
macroscopic condensate, with
 $N_0 {\ \lower-1.2pt\vbox{\hbox{\rlap{$<$}\lower5pt\vbox
{\hbox{$\sim$}}}}\ } N$ and only small
depletion
($N'/N\equiv 1 - N_0/N\ll 1$).   For a  repulsive scattering  length
$a >0$ and mean atomic density  $n$, this  depletion is of order
$\sqrt{na^3}$ \cite{MBT}
 and thus nonperturbative.

The balance between kinetic energy
$\hbar^2/2m\xi^2$ and interaction energy
$4\pi a \hbar^2n/m$ defines the ``healing length'' $\xi =(8\pi a
n)^{-1/2}$, and the Bogoliubov approximation
requires that $a
\ll
\xi$.  In addition, a confined Bose gas with typical dimension
$R$ differs qualitatively from  an ideal Bose gas whenever  $\xi \ll R$.
The present work uses the Bogoliubov
approximation to provide  a hydrodynamic description of the excited states
and  a variational estimate of the
eigenvalue
$\omega_l$
 and zero-temperature occupation number $N_l'$
for the lowest radial mode
 for each positive angular momentum $l\ge 1$.

\section{BASIC FORMALISM}
In \cite{And}, the harmonic trap $U =
\frac{1}{2}m\omega_0^2r^2$ (here taken as isotropic) has a size
$d_0=\sqrt{\hbar/m\omega_0} \approx 1~\rm \mu m$, and  the positive
scattering length $a \approx
10~\rm nm$ acts to expand  the  atomic cloud.   For  ${\cal N} \equiv
Na/d_0\gg 1$, the actual atomic
density
$\approx N/R^3$  is smaller than the naive estimate
$\approx N/d_0^3$, with the radial
expansion factor $R/d_0$  of order
${\cal N}^{1/5}$ \cite{BP}; thus the Bogoliubov approximation  holds for
$Na^3/R^3 \approx {\cal
N}^{2/5}(a/d_0)^2\ll 1$ [namely,
$N \ll (d_0/a)^6 \sim 10^{12}$]. The additional condition $R\gg \xi$ for
the failure of an ideal-Bose-gas
model now requires
$(8\pi {\cal N}^{4/5})^{1/2}
\gg 1$, which here holds even for $N \approx 100$.

The condensate wave function $\Psi$ satisfies the nonlinear
Gross-Pitaevskii equation \cite{G,P}
\begin{equation} \left(T + U - \mu +V\right)\Psi = 0,
\end{equation}
where $T = -\hbar^2\nabla^2/2m$ is the kinetic energy, $V \equiv 4\pi a
\hbar^2m^{-1}|\Psi|^2$ characterizes
the repulsive interaction energy, and
 the chemical potential $\mu$  must be adjusted to ensure that $\int dV
\,|\Psi|^2 = N_0$ and that $\Psi \to
0$ at large distances. In the present case, the condensate is stationary
with density
$n_0(r) = |\Psi(r)|^2$.  For large
${\cal N}
$, the kinetic energy can be ignored \cite{BP}, giving the ``Thomas-Fermi''
approximation
\begin{equation}
V_{TF} = (\mu - U)\,\theta(\mu - U)  \equiv
{\textstyle\frac{1}{2}}\hbar\omega_0(R^2 -  r^2)\,\theta(R -
r),\label{eq:TF}\end{equation}
where lengths are expressed in units of $d_0$,  $\mu = \frac{1}{2}
\hbar\omega_0 R^2$ defines the
dimensionless ``radius'' $R$ of the condensate, and  $R^5 =15{\cal N}$
\cite{BP}.

\section{HYDRODYNAMIC DESCRIPTION}
 The (small) noncondensate field operator
$\hat\phi$  obeys the linear equation
\begin{equation}i\hbar\partial \hat\phi/\partial t = \left(T+U - \mu +
2V\right)\hat\phi +
V\hat\phi^{\dagger},\label{eq:ex}\end{equation}
along with the adjoint equation.  In the Bogoliubov approximation ($N'\ll
N_0$), the total  particle density
operator becomes
$\hat\psi^{\dagger}\hat\psi\approx n_0 + \hat\rho'$, where the operator
$\hat\rho'=\sqrt{n_0}(\hat\phi +\hat\phi^{\dagger})$ characterizes the
fluctuating  noncondensate density;
similarly,  the particle current operator here becomes $\hat {\bf j}'
\approx n_0\nabla
\hat \Phi'$, where $\hat\Phi'= (\hbar/2mi\sqrt{n_0})(\hat\phi -
\hat\phi^{\dagger})$ is the velocity potential
operator.  Equation (\ref{eq:ex}) readily yields
%
\begin{equation}\partial\hat\rho'/\partial t +\nabla\cdot
(n_0\nabla\hat\Phi')  = 0,\label{eq:cont}\end{equation}
and the linearized Bernoulli's equation \cite{ALF}
\begin{equation}\frac{\partial\hat\Phi'}{\partial t} + \frac{4\pi
a\hbar^2}{m^2}\hat\rho'+
\frac{\hbar^2}{4m^2n_0}\left[\nabla\cdot\left(\hat\rho'\frac{\nabla
n_0}{n_0}\right)
-\nabla^2\hat\rho'\right] = 0.\label{eq:Bern}\end{equation}
The latter  can be recognized as a
linearization of the exact classical Bernoulli's theorem \cite{FW} $U
+{\textstyle\frac{1}{2}}mv^2 + (e+p)/n +
m\partial\Phi/\partial t = 0$ for a compressible irrotational isentropic
fluid with density
$n_0+\hat\rho'$  and velocity potential~$\hat\Phi'$
[here, $p(n) = 2\pi a \hbar^2n^2/m$  is the pressure functional and $e(n) =
\sqrt n\, T\sqrt n + p(n) $ is the
(quantum)  energy-density functional].

\section{BOGOLIUBOV EQUATIONS}
The hydrodynamic description provides  a valuable  qualitative picture of
the normal modes of
the nonuniform compressible spherical condensate, but the resulting
equations involve  spatial
derivatives of
$n_0$;  furthermore, in the Thomas-Fermi approximation (\ref{eq:TF}),  the
appropriate boundary conditions at
$R$  are not obvious.  Thus it is preferable to return to   Eq.~(\ref{eq:ex}),
expressing the field operator $\hat\phi$ as a linear combination of
quasiparticle operators $\alpha_j$ and
$\alpha_j^{\dagger}$
\begin{equation}\hat\phi({\bf r}, t) = \mathop{{\sum}'}_j\left[u_j({\bf
r})\alpha_je^{-i\omega_j t} -v_j^*({\bf
r})\alpha_j^{\dagger}e^{i\omega_j t}\right];\end{equation}
here, the wave-function coefficients $u_j$ and $v_j$ constitute a
two-component vector function ${\cal U}_j$ that
obeys the matrix ``Bogoliubov'' eigenvalue equation
\cite{ALF}
\begin{equation}(T + U -\mu + 2V)\,{\cal U}_j -V\,\tau_1\,{\cal U}_j =
\hbar\omega_j\,\tau_3\,{\cal
U}_j.\label{eq:Bog}\end{equation}
In contrast to Eqs.~(\ref{eq:cont})
and (\ref{eq:Bern}), only $V \propto n_0$  appears here, and the  quantum
interpretation provides the
obvious  boundary condition that ${\cal U}_j$ vanish for $r \to \infty$
(note that the hydrodynamic amplitudes
are essentially linear combinations of $u_j$ and $v_j$).

In the case of a static spherical condensate, the Bogoliubov
Eqs.~(\ref{eq:Bog}) have solutions of the form
$u_l(r)Y_{lm}$ and $v_l(r)Y_{lm}$;  in addition, they
 have a simple variational basis \cite{ALF} that yields an upper bound for
the lowest eigenvalue $\omega_l$ for
each positive angular momentum $l>0$.  To ensure the proper normalization,
take  $u_l(r) = \cosh \chi_l\,f_l(r)$
and
$v_l(r) =
\sinh
\chi_l\,f_l(r)$, where  $\int_0^{\infty}r^2dr\,|f_l(r)|^2 = 1$. With the
Thomas-Fermi approximation
(\ref{eq:TF}), the analogy to acoustic waves in a sphere  suggests  the
radial trial function
$f_l(r) \propto x^l(1-x^2)\,\theta(1-x)$, where $x \equiv r/R$.  A simple
variational calculation
yields the estimate $\omega_l/\omega_0 =  \sqrt{T_l^2 + 2T_lV_l}$, with
$T_l = \frac{1}{4}(2l+3)(2l+7)/R^2$ and
$V_l = 3R^2/(2l+9)$ the expectation values of the dimensionless kinetic
energy $-\frac{1}{2}\nabla^2$ and
Thomas-Fermi potential energy $\frac{1}{2}(R^2 - r^2)$.  This frequency has
a very different form for small and
large
$l$
\begin{equation}\frac{\omega_l}{\omega_0}\approx
\cases{\sqrt{\frac{3}{2}(2l+3)(2l+7)/(2l+9)},& for
$T_l \ll V_l$;\cr(2l+3)(2l+7)/4R^2,&for $T_l\gg V_l$.\cr}\end{equation}
 Stringari \cite{Str} has obtained a  related expression $\omega_l/\omega_0
= \sqrt l$ with a
purely hydrodynamic model, adding that $\omega_1 = \omega_0$ is an exact
result.   Note the close analogy to
the Bogoliubov excitation energy $E_k =
\sqrt{T_k^2 + 2T_kV_k}$ for a uniform bulk medium, where $T_k =
\hbar^2k^2/2m$ and $V_k = 4\pi a
\hbar^2n_0/m$ \cite{MBT}.

Each of these lowest excited
states for $l>0$ has a zero-temperature occupation number $N_l' =
\int_0^Rr^2dr\,|v_l(r)|^2$,  with the
variational estimate
$N_l' = \sinh^2\chi_l =  \frac{1}{2}(T_l~+~V_l)/\sqrt{T_l^2 + 2T_lV_l} -
\frac{1}{2}$, again very similar in
structure  to that  for a uniform Bose gas \cite{MBT}.  In particular,
$N_l' \approx \sqrt{V_l/8T_l}\gg 1$ for
$T_l \ll V_l$, and
$N_l'
\approx V_l^2/4T_l^2\ll 1$ for $T_l \gg V_l$.

\acknowledgements
Supported in part by the National Science
Foundation under Grant No. DMR 94-21888;  I am grateful for helpful
correspondence with B.~V.~Svistunov.

\end{document}